\documentclass[iop,letterpaper]{emulateapj}
\usepackage{multirow}

\newcommand{\dfplot}[1]{\plotone{figs/#1}}

\newcommand{\dfplottwo}[2]{\plottwo{figs/#1}{figs/#2}}

\newcommand{\D}[1][]{\ensuremath{\Delta_{#1}}}
\newcommand{\Teff}{\ensuremath{T_{\mathrm{eff}}}}
\newcommand{\logg}{\ensuremath{\log g}}
\newcommand{\logZ}{\ensuremath{\log Z}}
\newcommand{\mum}{~\ensuremath{\mu \mathrm{m}}}

\newcommand{\K}{\ensuremath{\mathrm{K}}}
\newcommand{\degree}{\ensuremath{^\circ}}
\newcommand{\EBVSFD}{\ensuremath{E(B-V)_\mathrm{SFD}}}
\newcommand{\SN}{\ensuremath{\mathrm{S}/\mathrm{N}}}
\newcommand{\AAm}{\text{\AA}}

\begin{document}

\title{Measuring Reddening with SDSS Stellar Spectra and Recalibrating SFD}

\author{Edward F. Schlafly\altaffilmark{1}, 
Douglas P. Finkbeiner\altaffilmark{1,2}
}

\altaffiltext{1}{Physics Department, Harvard University, Cambridge, MA 02138, USA}
\altaffiltext{2}{Harvard-Smithsonian Center for Astrophysics, 60 Garden St., Cambridge, MA 02138, USA}

\begin{abstract}
We present measurements of dust reddening using the colors of stars with spectra in the Sloan Digital Sky Survey.  We measure reddening as the difference between the measured and predicted colors of a star, as derived from stellar parameters from the SEGUE Stellar Parameter Pipeline \citep{Lee:2008a}.  We achieve uncertainties of 56, 34, 25, and 29 mmag in the colors $u-g$, $g-r$, $r-i$, and $i-z$, per star, though the uncertainty varies depending on the stellar type and the magnitude of the star.  The spectrum-based reddening measurements confirm our earlier ``blue tip'' reddening measurements \citep[S10]{Schlafly:2010}, finding reddening coefficients different by $-3$\%, 1\%, 1\%, and 2\% in $u-g$, $g-r$, $r-i$, and $i-z$ from those found by the blue tip method, after removing a 4\% normalization difference.  These results prefer an $R_V=3.1$ \citet[F99]{Fitzpatrick:1999} reddening law to \citet{O'Donnell:1994} or \citet{Cardelli:1989} reddening laws.  We provide a table of conversion coefficients from the \citet[SFD]{Schlegel:1998} maps of $E(B-V)$ to extinction in 88 bandpasses for 4 values of $R_V$, using this reddening law and the 14\% recalibration of SFD first reported by S10 and confirmed in this work.

\emph{Subject headings: }
dust, extinction --- 
Galaxy: stellar content --- 
ISM: clouds
\end{abstract}
\maketitle

\section{Introduction}

Dust is composed of heavy elements produced by the nuclear burning of stars.  These heavy elements are blown out of the stars in winds and explosions, and are reprocessed in the interstellar medium to eventually form dust grains \citep{Draine:2003}.  The dust scatters and absorbs light, especially in the ultraviolet through infrared, according to the dust reddening law.  The dust also emits photons thermally in the far-infrared.  Accordingly, mapping the dust is a central problem in astronomy.

In previous work with the blue tip of the stellar locus \citep[S10]{Schlafly:2010}, we examined the reddening law and the accuracy of the \citet[SFD]{Schlegel:1998} dust map using photometry from the Sloan Digital Sky Survey \citep[SDSS]{York:2000} and the uniformity of the color of the blue tip of the stellar locus over the sky.  This blue tip work recommended a 14\% recalibration of the SFD dust map in the sense that $E(B-V) = 0.86\cdot\EBVSFD$, and a preference for a \citet[F99]{Fitzpatrick:1999} reddening law over other reddening laws.  In this work we set out to test that result using an independent set of data.  

The SDSS stellar spectra provide an independent test of reddening.  Stellar spectra sensitively test reddening because the broadband photometry of a star is almost completely determined by three parameters: temperature, metallicity, and gravity.  These paramaters can be determined using only line information in the spectra, allowing the intrinsic broadband colors of the star to be predicted independently from the observed colors of the star.  Dust intervening between us and the star, however, will shift the observed colors relative to the intrinsic colors.  The difference between the predicted intrinsic colors and measured colors constitutes a measurement of the reddening to that star.  This method is broadly similar to that of \citet{Peek:2010} and \citet{Jones:2011}, which use SDSS galaxy and M-dwarf spectra, respectively, to similar effect.

The eighth data release of the SDSS has spectra for about 500,000 stars \citep{Aihara:2011}.  The SEGUE Stellar Parameter Pipeline \citep[SSPP]{Lee:2008a} has uniformly processed these spectra to measure the temperature, metallicity, and gravity of each of these stars using a variety of methods, including ones that are independent of the observed colors of the star.  We use one such method to predict the intrinsic colors of each star.  The difference between the observed colors and intrinsic colors is used as a measurement of reddening to each star.  These reddening measurements are then used to test the calibration of SFD and the reddening law.

We test the calibration of SFD and the reddening law by comparing our reddening measurements with the predictions from SFD and a reddening law.  A reddening law predicts, in each color $a-b$, the reddening $E(a-b)$ relative to the reddening in some reference color, usually $B-V$.  The SFD dust map predicts $E(B-V)$, effectively giving the normalization of the reddening law in any direction on the sky.  We use our reddening measurements to find $R_{a-b} = E(a-b)/\EBVSFD$ for each of the SDSS colors, testing both the reddening law and the SFD normalization.  We also find the ratios of the $R_{a-b}$, which test the reddening law without using SFD as a reference.

This work additionally lays the foundation for future investigations of the three-dimensional distribution of dust.  Because we have stellar parameters for each of the stars, we can obtain accurate absolute magnitudes and distances.  The SDSS targets both blue-horizontal-branch (BHB) stars and M-dwarf stars, in principle permitting the dust to be studied over a wide range of distances.  In this work, however, we have focused on the two-dimensional distribution of the dust.

The paper is organized as follows: in \textsection \ref{sec:data}, we present the data sets used in this work.  In \textsection \ref{sec:methods}, we describe our method for transforming SSPP stellar parameters and SDSS photometry into reddening measurements.  In \textsection \ref{sec:results}, the reddening measurements are presented, calibrated, and mapped.  In \textsection \ref{sec:discussion} and \textsection \ref{sec:conclusion}, we discuss the implications of these results, especially for the reddening law, and conclude.

\section{Data}
\label{sec:data}

To measure the reddening to a star, we require its observed colors and a prediction for its intrinsic colors, as derived from stellar parameters for that star.  The SDSS provides all of this information.  In this section, we first describe the SDSS imaging, which provides the observed colors, and the SDSS spectroscopy and SSPP, which provide the stellar parameters.  We then present the MARCS grid of model atmospheres \citep{Gustafsson:2008}, which is used to connect the stellar parameters to colors.  Finally, we describe the cuts we perform on the full set of SDSS stars with spectra to get the sample of stars used in this work.

\subsection{The SDSS}
\label{subsec:sdss}
The eighth data release of the SDSS provides uniform, contiguous imaging of about one third of the sky, mostly at high Galactic latitudes \citep{Aihara:2011}.  The SDSS imaging is performed nearly simultaneously in five optical filters: $u$, $g$, $r$, $i$, and $z$ \citep{Gunn:1998, Fukugita:1996}.  The photometric pipeline has uniformly reduced data for about $10^8$ stars.  The SDSS is 95\% complete up to magnitudes 22.1, 22.4, 22.1, 21.2, and 20.3 in the bandpasses $u$, $g$, $r$, $i$, and $z$, which have central wavelengths of about 3600, 4700, 6200, 7500, and 8900~\AA.   We use SDSS data that have been photometrically calibrated according to the ``ubercalibration'' procedure of \citet{Padmanabhan:2008}.

The technique presented here is sensitive to the SDSS photometric calibration, and so we briefly describe the basic units in which the SDSS data is calibrated: the run and camera column (``camcol'').  The SDSS takes imaging data in runs that usually last for several hours.  These runs observe regions of sky that are long in right ascension and narrow in declination.  Each run is composed of observations from six camcols: the six sets of five CCDs (one for each of the photometric bands) that fill the SDSS focal plane.

The SDSS photometry is used to select targets for follow-up SDSS spectroscopy.  Spectroscopy is performed with two multiobject double spectrographs, which are fed by 640 fibers.  The spectra cover a wavelength range of $3800-9200\AAm$ at a resolution of $\lambda/\Delta \lambda \simeq 2000$.   The stars used in this work were targeted for spectroscopy either as part of the main SDSS survey or as part of the Sloan Extension for Galactic Understanding and Exploration \citep[SEGUE]{Yanny:2009} and its successor, SEGUE-2.  Targets from the main survey are observed until they reach a target signal to noise ($\SN$) of about $(\SN)^2 > 15/\mathrm{pix}$ for stars with magnitudes $g = 20.2$, $r = 20.25$, $i = 19.9$.  SEGUE targets are exposed long enough to achieve $(\SN)^2 > 100$ at the same depth, so that stellar parameters can be measured \citep{Abazajian:2009}.

\begin{figure}[tbh]
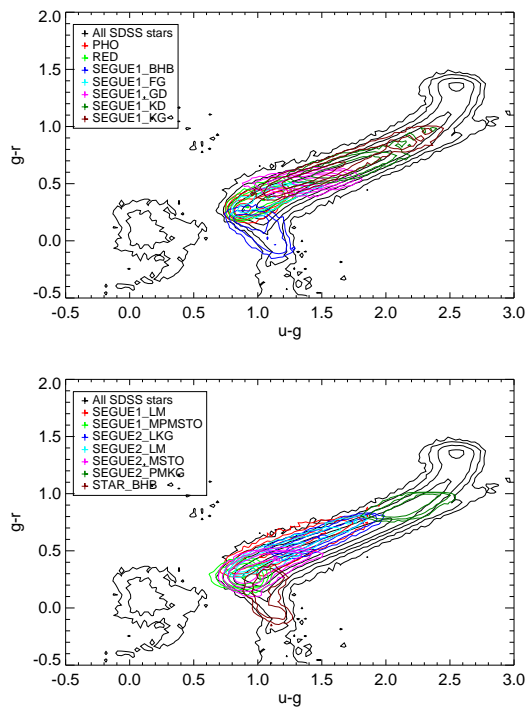

\dfplot{seguesl.ps}
\figcaption{
\label{fig:seguesl}
Colors of stars with SDSS spectra used in this work.  The black contours give the number density of all point sources detected by the SDSS at greater than 20$\sigma$ confidence in the $ugri$ bands.  The colored contours give the number densities of stars with SDSS spectra, targeted according to the program labeled in the legend.  In each panel the horizontal axis gives $u-g$ and the vertical axis gives $g-r$, in mags; the figure is split to make the contours more legible.
}
\end{figure}

The SDSS has a number of programs targeting different stellar types for spectra.  Figure~\ref{fig:seguesl} shows a $ugr$ color-color plot of the different types of stars used in this work, together with the overall stellar locus.  \citet{Yanny:2009} gives a detailed description of the criteria that define these targets.  The stars used in this work cover most of the stellar locus, except for the reddest stars, for which stellar parameters are poorly measured.  Table~\ref{tab:targettypes} gives the names of each of the programs targeting stars used in this analysis, as well as the number of stars used.  In addition to stars targeted for specific science goals, we also use the reddening and spectrophotometric standard star target types.  These stars are F stars used for calibration purposes in the SDSS spectroscopic pipeline.

\tabletypesize{\scriptsize}
\begin{deluxetable}{c c c}
\tablewidth{\columnwidth}
\tablecaption{SDSS spectral targets}
\tablehead{
\colhead{Target type} & \colhead{\# of stars} & \colhead{Description}
}
\startdata
            PHO &  15894 & Spectrophotometric standards \\
            RED &  14353 & Reddening standards \\
    SEGUE1\_BHB &  12603 & BHB stars \\
     SEGUE1\_FG &   5065 & FG stars \\
     SEGUE1\_GD &  44495 & G dwarfs \\
     SEGUE1\_KD &  13459 & K dwarfs \\
     SEGUE1\_KG &  16012 & K giants \\
     SEGUE1\_LM &  22273 & low metallicity stars \\
SEGUE1\_LOW\_KG &   2606 & low-latitude K giants \\
SEGUE1\_LOW\_TO &   6432 & low-latitude turn-off stars \\
 SEGUE1\_MPMSTO &  26885 & metal-poor F stars \\
    SEGUE2\_LKG &  19634 & K giants \\
     SEGUE2\_LM &  13785 & low metallicity stars \\
   SEGUE2\_MSTO &  31767 & main sequence turn-off stars \\
   SEGUE2\_PMKG &  10566 & K giants \\
      STAR\_BHB &   5667 & Main survey BHB stars \\
          Total & 261496 &
\enddata
\tablecomments{
\label{tab:targettypes}
The target types and the number of stars used for stars considered in this work.  The low-latitude targets are selected according to targeting algorithms designed to select stars in highly-reddened low Galactic latitude regions.  SEGUE-1 \citep{Yanny:2009} and SEGUE-2 provide most of the targets for this work.  The main survey, however, provides the STAR\_BHB targets, as well as most of the spectrophometric and reddening standards.
}
\end{deluxetable}
\tabletypesize{\footnotesize}

\subsection{The SSPP}
\label{subsec:sspp}

The Sloan Extension for Galactic Understanding and Exploration is a spectroscopic and photometric extension to the SDSS designed to provide insight into the structure and history of the Galaxy \citep{Yanny:2009}.  As part of this program, the SSPP was developed to uniformly process the stellar spectra from the SDSS \citep{Lee:2008a}.  This work uses the eighth data release of the SDSS, which includes an updated version of the SSPP, and spectra from SEGUE-2 in addition to SEGUE.  The SSPP provides multiple estimates of the spectral type, metallicity, gravity, and radial velocity for all of the stars with SDSS spectra.  The accuracy of these parameters has been repeatedly tested by comparison with high resolution spectroscopy and by using measurements from globular and open clusters \citep{Lee:2008b, AllendePrieto:2008, Smolinski:2010}.

The SSPP provides a number of different estimates for the temperature, metallicity, and gravity of each star, and a single composite estimate.  We cannot use the composite estimates produced by the SSPP because some of the SSPP parameter estimates rely on the photometry of the stars, which is affected by the dust.  Instead we rely only on methods that do not make use of any photometry.  These are the \texttt{NGS1}, \texttt{ki13}, \texttt{ANNRR}, and \texttt{ANNSR} methods \citep{Lee:2008a}.  For this work we have used the \texttt{ANNRR} estimator \citep{ReFiorentin:2007}, which uses a neural net trained on continuum-normalized SDSS spectra to make parameter estimates.  However, the four methods produce compatible stellar parameters and the results of this work are insensitive to the method used.

\subsection{MARCS Model Atmospheres}
\label{subsec:grid}

The stellar parameters derived from the SSPP are transformed into predicted $ugriz$ colors using the MARCS grid of model atmospheres \citep{Gustafsson:2008}.  Each model includes estimates of the surface flux at a resolution of $\lambda/\Delta \lambda$ of 20,000, and covers the range 1,300--200,000~\AA.  The grid covers an extensive range of stellar parameters: $2500~\K \le T_\mathrm{eff} \le 8000~\K$, $-5 \le [\mathrm{Fe}/\mathrm{H}] \le 1$, and $-1 \le \logg \le 5.5$, as well as additional parameter combinations involving $\alpha$-enhancement and microturbulence.  \citet{Edvardsson:2008} and \citet{Plez:2008} have verified the accuracy of synthetic broadband colors predicted from this grid.  For $\logg < 3$, we use the MARCS spherical stellar atmosphere models, while for $\logg \ge 3$, we use the plane parallel models.  We have experimented with using other grids: the \citet{Munari:2005} grid, the \citet{Castelli:2004} grid, and the NGS1 grid used internally in the SSPP.  The \citet{Munari:2005} grid predicts bluer colors than the other grids owing to the grid's not including ``predicted lines,'' faint lines which have not been individually detected \citep{Munari:2005}.  Nevertheless, because we calibrate the predicted colors to match the observed colors (\textsection \ref{subsec:calibration}), our final results are insensitive to the choice of grid, even when the raw predicted colors are substantially discrepant from the observed colors.

\subsection{Selection Cuts}
\label{subsec:cuts}

The SDSS targets a wide variety of stars for spectroscopy, including exotic types for which the spectral parameters are difficult to measure.  We are interested in the dust, and so want to select stars for which the stellar parameters are well understood and accurately measured.  Accordingly, we exclude from this analysis stars targeted to have unusual colors and very red stars with dereddened $g-r > 1$ mag.  Likewise, we exclude any white dwarf targets.  Finally, we exclude any objects with stellar parameters marked as unreliable by the SSPP (according the parameters' indicator variables).

The remaining target types and number of stars used in each type are listed in Table~\ref{tab:targettypes}.  We find it convenient to occasionally divide these target types into five classes.  These classes are the standard stars, which were targeted as reddening standards or spectrophotometric standards; the FG stars, targeted to have spectral types F or G; the BHB stars; the K stars, targeted to have spectral near K; and the other stars, which include the low-metallicity and low-latitude targets.

We have chosen not to impose a cut on the $\SN$ of the spectra.  We have varied cuts on $\SN$ from 0 to 50; the resulting best fit reddening coefficients presented in this work vary by only 0.5\% in that range.

The total number of stars used following these cuts is 261,496.

\section{Methods}
\label{sec:methods}

We predict the colors of stars using the stellar parameters from the SSPP and the MARCS grid.  We compare these predicted colors to the measured colors from the SDSS imaging to derive the reddening to each star.

The predicted broadband magnitudes $m_b$ are given by integrating each synthetic spectrum over the SDSS system throughput for each of the five bands $b$ \citep{Gunn:1998}, according to
\begin{equation}
m_b = -2.5 \log \frac{\int d\lambda S(\lambda) W_b(\lambda)}{\int d\lambda Z(\lambda) W_b(\lambda)}
\end{equation}
Here the source spectrum $S$ has units photons/s/\AA, the system throughput $W_b$ is unitless, and the AB magnitude reference spectrum $Z$ has the same units as $S$.  The AB magnitude reference spectrum $Z$ is simply a flat spectrum with $F_\nu = 3631\ \mathrm{Jy}$ \citep{Oke:1983}.

The magnitude of a source in a single band predicted in this way is not useful because it depends on the unknown radius and distance of the source.  However, in this work we consider only colors, for which the distance and radius dependence cancels.  We could in principle use the stellar parameters and stellar evolutionary tracks to determine the radius of the star, and so derive distances in conjunction with reddenings, but we defer this to future work.

By computing the magnitudes corresponding to each synthetic spectrum, we construct a synthetic grid of magnitudes.  The magnitudes for each star are then predicted by linearly interpolating off this grid.  The differences between the measured colors and predicted colors give reddening estimates.  We define this reddening measurement in the color $a-b$ to be $\D[a-b] = (a-b)_\mathrm{obs} - (a-b)_\mathrm{pre}$.

\section{Results}
\label{sec:results}

The reddening estimates $\D[a-b]$ are used to measure the dust reddening law and test dust properties.  In this section, we describe the performance of these reddening measurements at high Galactic latitudes (\textsection \ref{subsec:highlat}), demonstrating the need for an empirical calibration to improve the measurements.  We then perform that calibration (\textsection \ref{subsec:calibration}).  Finally, we present maps of the calibrated reddening measurements (\textsection \ref{subsec:redmap}).

\subsection{Performance at High Galactic Latitudes}
\label{subsec:highlat}

We test the quality of the $\D$ by restricting to the region of sky with $|b| > 50\degree$ and $\EBVSFD < 0.04$.  In this region of low extinction, excursions of $\D$ from zero should be due primarily to the statistical uncertainties in the photometry and stellar parameters.  However, additional systematic effects can cause $\D \neq 0$.  These include any mismatch between the synthetic spectra and the real spectra and biases in the SSPP-derived stellar parameters.

The performance of $\D$ for this sample of stars is given in Table~\ref{tab:cal}, for several classes of star.  The rows marked ``raw'' give the means and standard deviations of the uncalibrated $\D$ relevant here.  In all colors, the mean $\D$ is within about 30 mmag of 0, comparable with the 10--20 mmag uncertainty in the SDSS zero points \citep{Abazajian:2004}.  In $r-i$ and $i-z$, the scatter in $\D$ is 20--30 mmag, comparable to the SDSS photometric uncertainty.

\begin{deluxetable*}{c c c c c c c c c}
\tablewidth{0pc}
\tablecaption{Bias in Predicted Magnitudes}
\tablehead{
\colhead{Target class} & \colhead{$\overline{\D[u-g]}$} & \colhead{$\sigma_{u-g}$} & \colhead{$\overline{\D[g-r]}$} & \colhead{$\sigma_{g-r}$} & \colhead{$\overline{\D[r-i]}$} & \colhead{$\sigma_{r-i}$} & \colhead{$\overline{\D[i-z]}$} & \colhead{$\sigma_{i-z}$}
}
\startdata
Standards (raw) &     22 &     51 &      3 &     33 &     28 &     22 &     17 &     25 \\ 
Standards (cal) &     -3 &     46 &     -2 &     31 &     -1 &     22 &     -1 &     24 \\ 
       FG (raw) &      0 &     82 &     -2 &     42 &     30 &     28 &     20 &     35 \\ 
       FG (cal) &      1 &     72 &     -2 &     37 &      0 &     27 &      0 &     34 \\ 
      BHB (raw) &     64 &     83 &     -1 &     50 &     36 &     32 &     33 &     42 \\ 
      BHB (cal) &     24 &     90 &    -13 &     46 &      0 &     31 &     -2 &     40 \\ 
        K (raw) &     10 &    114 &    -24 &     57 &     29 &     32 &     14 &     32 \\ 
        K (cal) &      5 &     73 &      4 &     35 &      0 &     27 &      0 &     28 \\ 
    Other (raw) &     13 &     88 &     20 &     47 &     34 &     28 &     18 &     31 \\ 
    Other (cal) &    -16 &     68 &      8 &     38 &      3 &     27 &      0 &     30 \\ 
      All (raw) &     12 &     92 &     -6 &     51 &     30 &     29 &     18 &     33 \\ 
      All (cal) &      1 &     71 &      0 &     37 &      0 &     27 &      0 &     31
\enddata
\tablecomments{
\label{tab:cal}
The mean ($\overline{\D}$) and standard deviation ($\sigma$) of $\D$ in mmag for different types of stars (\textsection \ref{subsec:sspp}), before and after calibration (\textsection \ref{subsec:calibration}), in each of the SDSS colors.  Rows marked ``(raw)'' or ``(cal)'' indicate whether the means and standard deviations are given before or after calibration, respectively.  After calibration, the mean of $\D$ is close to zero, and the scatter in $\D$ is reduced, especially in the color $u-g$.
}
\end{deluxetable*}

\begin{figure*}[!tbh]
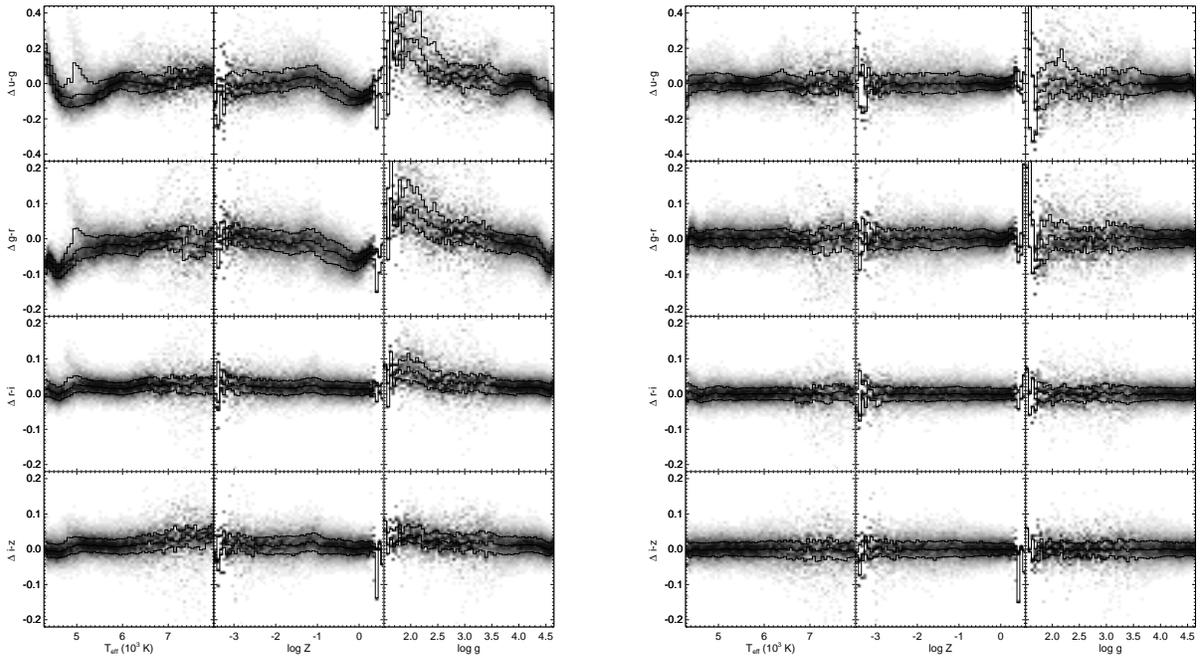

\dfplottwo{trendsraw.ps}{trendscal.ps}
\figcaption{
\label{fig:trends}
Color difference $\D$ in mags with temperature $\Teff$, metallicity $\logZ$, and gravity $\logg$ for the colors $u-g$, $g-r$, $r-i$, and $i-z$.  The left hand panels show $\D$ before calibration, while the right hand panels show $\D$ after calibration.  Before calibration, the $u-g$ color is worst predicted, and small but easily detectable trends with temperature exist for all of the colors.  Note the different scale for $u-g$.
}
\end{figure*}

However, in the colors $u-g$ and $g-r$, the scatter in $\D$ is somewhat larger than expected.  The expected uncertainty in $\D[u-g]$ and $\D[g-r]$ is about 40 and 30 mmag, respectively, including both the photometric uncertainty and the uncertainty in the stellar parameters.  However, the scatter in $\D$ in this sample of stars is 92 and 51 mmag in the colors $u-g$ and $g-r$.  Figure~\ref{fig:trends} (left panels) shows that additionally there are trends in $\D$ with temperature, metallicity, and gravity, especially in the color $u-g$.  Because we expect that typical stellar metallicity is correlated with the dust column to that star, these trends could bias this analysis and must be removed.  Moreover, these trends contribute to the scatter in $\D[u-g]$ and $\D[g-r]$.

We do not fully understand the source of these trends.  The two most likely candidates are mismatches between the synthetic stellar spectra and real stellar spectra, and biases in the SSPP stellar parameters.  We have tested for mismatch between the synthetic and observed spectra by comparing with other synthetic grids.  Comparing the grids of \citet{Gustafsson:2008}, \citet{Castelli:2004}, and \citet{Lee:2008a} indicates that the grids can disagree with one another by about 50 mmag in $u-g$ and $g-r$.  The larger effect, however, is probably due to biases in the SSPP stellar parameters.  The data release 8 version of the SSPP introduced substantial changes to metallicity estimates for stars around solar metallicity, by about 0.4 dex \citep{Smolinski:2010}.  Colors predicted using parameters from the data release 7 version of the SSPP were too blue by as much as 200 mmag in $u-g$.  

However, we calibrate the predicted colors to remove the trends and to render our results insensitive to these sources of systematic errors.  We have repeated the full analysis using four different grids \citep{Gustafsson:2008, Castelli:2004, Lee:2008a, Munari:2005}.  The final results of this work were unchanged at the 1\% level.  We have also repeated the analysis using the data release 7 and 8 versions of the SSPP, which changed the final results by at most 2\%.

\subsection{Calibration}
\label{subsec:calibration}

We remove these trends by calibrating the synthetic photometry so that $\D \approx 0$ on the high-latitude sample of stars where reddening is unimportant.  This procedure has been done before; for instance, \citet{Fitzpatrick:2005} also calibrate synthetic photometry to match observed photometry.  In this work, we perform the calibration by fitting an empirical curve to $\D$ as a function of temperature, metallicity, and gravity, and adjusting the synthetic photometry to remove this curve from $\D$.  

To properly model the trends in $\D$, we should find the function $f$ of temperature $\Teff$, metallicity $\logZ$, and gravity $\logg$ that most accurately predicts the colors.  To be more precise, $f$ should be chosen to minimize the distance between $\D$ and $f$ in color space, considering the covariance matrix for the observed colors and for the predicted colors.  We however deemed this computationally intractable and unnecessary, given the wide range of temperatures ($4000\ \K < \Teff < 8000\ \K$), metallicities ($-3 < \logZ < 0$), and gravities ($2 < \logg < 5$) available and the small estimated uncertainties in the stellar parameters for the objects we look at ($\sigma_T \lesssim 150 \K$, $\sigma_{\logZ} \lesssim 0.1$ dex, $\sigma_{\logg} \lesssim 0.1$ dex).  We therefore ignore the covariance between the stellar parameters and $\D$, and rely on the fact that the stars cover a broad range of parameters (much broader than the uncertainties in those parameters) to guarantee that the bias introduced by this is small.

To maximize the effective range of stellar parameters, we need to a set of stars for calibration that cover parameter space evenly without overweighting stars of a particular type.  Accordingly we choose stars from the $ugr$ color plane, capping the maximum number of stars chosen from any location in the plane.  This guarantees that all of the available stars are used in areas of color space where few stars were targeted, but limits the number of stars used in areas where many stars were targeted.  The primary consequence of this selection is to reduce the importance of the many SEGUE1\_GD targets chosen from a narrow range in temperature.

With this set of stars in hand, we find the function $f$ that minimizes 
\begin{equation}
\sum_i \left ( \frac{\D[c,i]-f_c(x_i)}{\sigma_{c,i}} \right )^2
\end{equation}
for each color $c$, where $i$ indexes over stars, $\D$ gives the measured color minus the predicted color, and $x$ gives the stellar parameters for the stars.  The uncertainty $\sigma$ is computed from the photometric uncertainty and the uncertainty in the predicted colors derived from the uncertainty in the stellar parameters.  The values $\D$ here are corrected for reddening according to S10, though as $\EBVSFD$ is less than 0.04 for this sample this is of little importance.  We use for $f$ a fifth order polynomial, though we have also used a second order polynomial with negligible effect on the final results of this work.

Figure~\ref{fig:trends} (right panels) shows that the remaining trends in $\D$ are small.  The rows in Table~\ref{tab:cal} marked ``cal'' show the means and standard deviations of $\D$ after calibration for several classes of star.  The calibration brings the mean of $\D$ to zero.  The calibration has the additional benefit of reducing the scatter in $\D$ to be much closer to the expected statistical uncertainty of about 40, 30, 25 and 30 mmag in $u-g$, $g-r$, $r-i$, and $i-z$.  

To verify that the final results are insensitive to the details of the calibration, we have varied the order of the polynomial used and the sampling method used to select calibration objects.  As long as the order of the polynomial is at least two, we achieve consistent results at the 1\% level.  The sampling method is similarly unimportant except in extreme cases where the calibration stars do not cover a satisfactory range of stellar parameters.

\subsection{Reddening Map}
\label{subsec:redmap}

\begin{figure*}[!tbh]
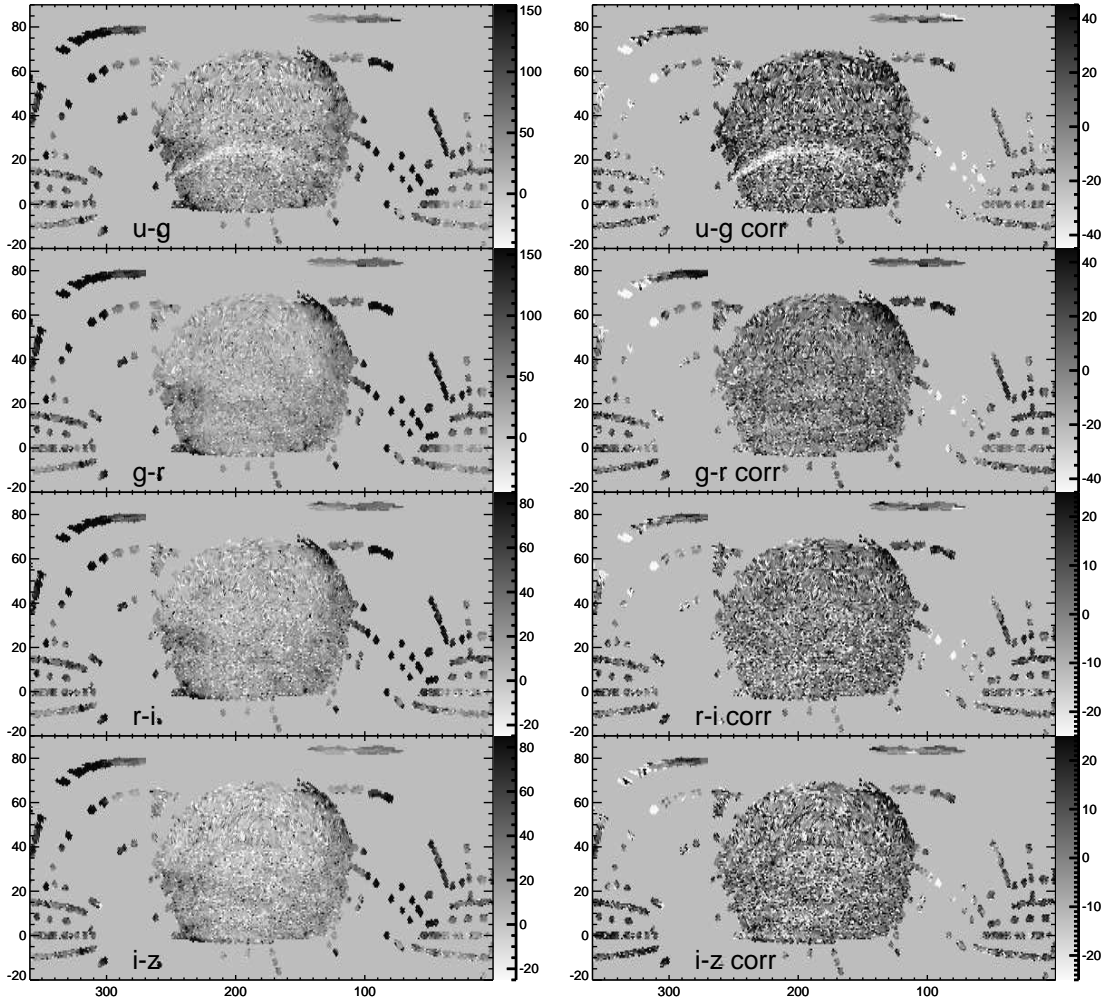

\dfplot{redmap.ps}
\figcaption{
\label{fig:redmap}
Map of $\D$ in $u-g$, $g-r$, $r-i$, and $i-z$ (left panels), giving the derived reddening to each star targeted by the SDSS.  The right hand panels (marked ``corr'') give the same maps, after correction for the dust according to SFD and the coefficients $R_{a-b}$ from \citet{Schlafly:2010}.  The color bars give the range of $\D$ in mmag corresponding to the grayscale figure.  The horizontal axis is right ascension and the vertical axis is declination, both in degrees.
}
\end{figure*}

The calibrated measurements of $\D$ immediately permit the construction of a reddening map, which is shown in Figure~\ref{fig:redmap} (left panels).  The right panels show the map after correction for dust according to SFD, using the prescription of S10.  The derived reddening map agrees well with that found in S10, clearly identifying the same runs with bad zero points in $u-g$ and the SFD-underpredicted region in the northwest of the north Galactic cap.  In the Galactic plane, the stars used for the reddening measurements may not be behind the entire dust column, and so it is unsurprising that their reddening is less than that predicted by SFD.  Over most of the observed sky, however, the residuals after correction for dust are within the uncertainties.

As pointed out in S10, we again find that the dereddened south is somewhat redder than the dereddened north (\textsection \ref{subsec:nsdiff}).  Apart from this trend and the cloud in the northwest, the residual maps are reassuringly flat and reveal no trends in color with Galactic latitude or longitude.

\begin{figure}[!tbh]
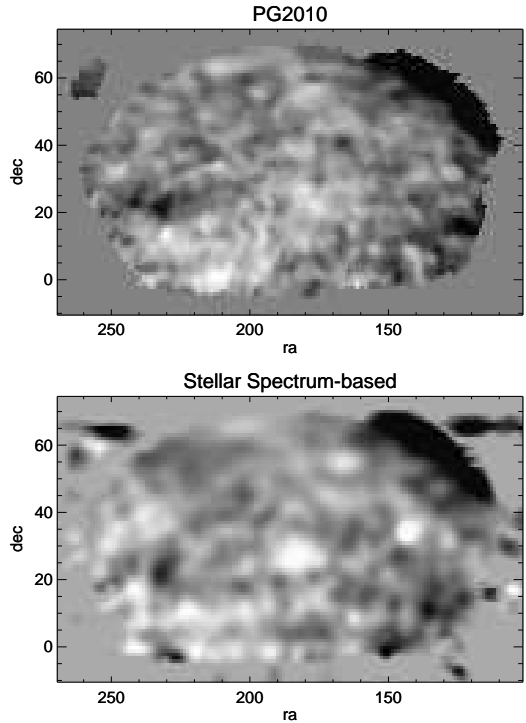

\dfplot{pgmap.ps}
\figcaption{
\label{fig:pgmap}
Color maps of the North Galactic Cap in $g-r$ after dereddening, from \citet{Peek:2010} and the stellar spectrum-based measurements of this work, smoothed to 4.5\degree.  Some of the same structures are visible.  The SFD underpredicted region in the northwest is obvious, but likewise in both sets of reddening measurements the southwest and northeast are redder than the southeast.  Both maps are scaled so that black is $-10$ mmag and white is $10$ mmag.  Axes give coordinates in right ascension and declination, in degrees.
}
\end{figure}

Figure~\ref{fig:pgmap} shows the residual $g-r$ map of \citet{Peek:2010}.  This map is compared with our residual $g-r$ map, smoothed to match the $4.5$ degree resolution of the \citet{Peek:2010} map.   The two maps show several consistent large scale features.  Both maps detect the SFD underprediction in the northwest, and additionally find that the southwest of the North Galactic Cap is somewhat redder than the southeast, by about 10 mmags.

\section{Discussion}
\label{sec:discussion}

The reddening measurements $\D$ permit tests of the reddening law and of the normalization of SFD.  In this section we use $\D$ to measure reddening in the SDSS colors, simultaneously constraining the normalization of SFD and the reddening law, using two methods.  In \textsection \ref{subsec:rab}, we find the best fit values of $R_{a-b}$, the reddening per unit $\EBVSFD$ in each of the SDSS colors.  In \textsection \ref{subsec:rabcc}, we find the best fit values of the ratios of the $R_{a-b}$, without using SFD.  In \textsection \ref{subsec:bias}, we discuss two possible sources of systematic error in the measurements: photometric bias from target selection and the bias introduced by including stars at different distances.  In \textsection \ref{subsec:regions} and \textsection \ref{subsec:nsdiff}, we examine the variation of reddening in different regions of the sky and a curious difference in residuals between the Galactic north and south.

\subsection{Fitting the Reddening in the SDSS Colors}
\label{subsec:rab}

We use the reddenings $\D$ to constrain $R_{a-b}$, the reddening in the color $a-b$ divided by \EBVSFD.  We measure the $R_{a-b}$ by fitting $\D$ as a linear function of $\EBVSFD$, the slope of which gives $R_{a-b}$.  We find that we can detect calibration errors in the SDSS using this technique.  Accordingly, we perform a second similar linear fit, where additionally a constant zero point offset in each SDSS run is also fit.

The $R_{a-b}$ are given by the equation
\begin{equation}
\label{eq:1}
\D[a-b] = R_{a-b} \EBVSFD + C
\end{equation}
where C gives a constant offset which should give zero if the calibration of the reddening values in the high-Galactic-latitude, low-reddening is accurate when extended over the entire footprint.  We find $R_{a-b}$ through least-squares minimization of 
\begin{equation}
\label{eq:2}
\chi^2 = \sum_i \left ( \frac{\D[a-b,i] - R_{a-b} E(B-V)_{\mathrm{SFD},i} - C}{\sigma_i} \right)^2
\end{equation}
where $i$ indexes over stars, $R_{a-b}$ and $C$ are the fit parameters, $E(B-V)_{\mathrm{SFD},i}$ is $\EBVSFD$ in the direction of star $i$, and $\sigma_i$ is the uncertainty in $\D[a-b,i]$, considering both the uncertainty in the measured color and the uncertainty in the predicted color, propagated from the uncertainty in the stellar parameters.

To render the fit insensitive to calibration errors in the SDSS, we also fit
\begin{equation}
\D[a-b] = R_{a-b} \EBVSFD + C_r
\end{equation}
in an analogous least-squares sense, where $r$ indexes over SDSS run number, which absorbs any zero point calibration errors in the SDSS runs.  We call this the fit with zero point offsets.  The results of this fit are similar to those given by Equation~\ref{eq:1}.  Each fit is iterated and clipped at $3\sigma$ in each color, until finally all clipped points are removed and the fit is repeated with the same clipped set of stars in each color.

The results from the fits are given in Table~\ref{tab:fitebv}.  The fits with and without zero point offsets agree to within 4\%.  Both sets of fits give values consistently slightly smaller than the S10 values ($\sim 4\%$).  In S10, we found that the best fit normalization of SFD varied over the sky by about 10\%, and so this 4\% normalization difference may be due to the different sky footprint available to the S10 and spectrum-based analyses.  However, when we tested this at high Galactic latitudes in the north where the two footprints overlapped, the results were inconclusive because the uncertainties were similar in size to the 4\% effect we observe.  The $\chi^2$ per degree of freedom ($\chi^2/\mathrm{dof}$) for the fits are quite good, especially in the redder colors; without including zero point offsets, we achieve 1.45, 1.19, 1.04, and 1.02 $\chi^2/\mathrm{dof}$ in the colors $u-g$, $g-r$, $r-i$, and $i-z$, respectively.

\tabletypesize{\scriptsize}
\begin{deluxetable}{c c c c c c c}
\tablewidth{\columnwidth}
\tablecaption{$R_{a-b}$ derived by fitting SFD}
\tablehead{
\colhead{Color} & \colhead{$R_{a-b}$} & \colhead{$\chi^2/N$} & \colhead{$R_{a-b}$ (zp)} & \colhead{$\chi^2_{\mathrm{zp}}/N$} & \colhead{S10} & \colhead{SFD}
}
\startdata
$u-g$ &  $0.94 \pm  0.02$ &  1.45 &  $0.98 \pm  0.01$ &  1.41 &  1.01 &  1.36 \\ 
$g-r$ &  $0.98 \pm  0.02$ &  1.19 &  $0.94 \pm  0.01$ &  1.12 &  1.00 &  1.04 \\ 
$r-i$ &  $0.55 \pm  0.01$ &  1.04 &  $0.56 \pm  0.01$ &  1.01 &  0.57 &  0.66 \\ 
$i-z$ &  $0.44 \pm  0.01$ &  1.02 &  $0.42 \pm  0.01$ &  0.97 &  0.45 &  0.61 \\ 
\enddata
\tablecomments{
\label{tab:fitebv}
$R_{a-b}$ derived from fitting SFD.  The column $R_{a-b}$ gives the fit results without zero point offsets, while the column $R_{a-b}$ (zp) gives the fit results with zero point offsets.  The $\chi^2$ per degree of freedom ($\chi^2/N$) is close to unity in the redder bands, and improves when zero points are fit ($\chi^2_{\mathrm{zp}}/N$).  The predictions for $R_{a-b}$ according to \citet[S10]{Schlafly:2010} and the combination of SFD and \citet{O'Donnell:1994} are also given.  The spectrum-based $R_{a-b}$ are seen to closely match one another and the S10 values, while disagreeing with the prediction according to SFD and \citet{O'Donnell:1994}.
}
\end{deluxetable}
\tabletypesize{\footnotesize}

These results confirm the preference for an $R_V=3.1$ F99 reddening law and a 14\% recalibration of SFD, first reported in S10.  The predicted $R_{a-b}$ according to SFD and an \citet{O'Donnell:1994} reddening law are in disagreement with the measurements.

\begin{figure}[tbh]
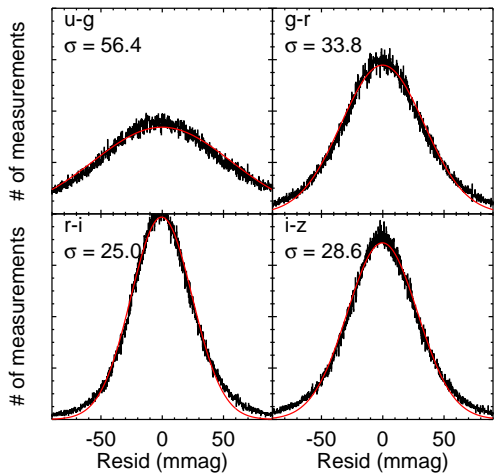

\dfplot{resid.ps}
\figcaption{
\label{fig:resid}
The distribution of residuals $\chi$ (Equation \ref{eq:2}) in mmag for the fit without zero point offsets.  The residuals are largely Gaussian, though the wings are non-Gaussian and are probably caused by errors in SFD and the changing best-fit dust normalization.  A Gaussian fit is overplotted, with standard deviation given by $\sigma$ in each panel.
}
\end{figure}

Figure~\ref{fig:resid} shows the residuals to the fit without zero point offsets.  The residuals are largely Gaussian, with $\sigma$ of 56, 34, 25, and 29 mmag in $u-g$, $g-r$, $r-i$, and $i-z$.  This scatter is substantially better than the scatter we achieved in the calibration sample of stars (Table~\ref{tab:cal}), because those stars were chosen to be widely distributed over the $ugr$ color-color plane and therefore many of those stars had outlying photometry.

\begin{figure*}[tbh]
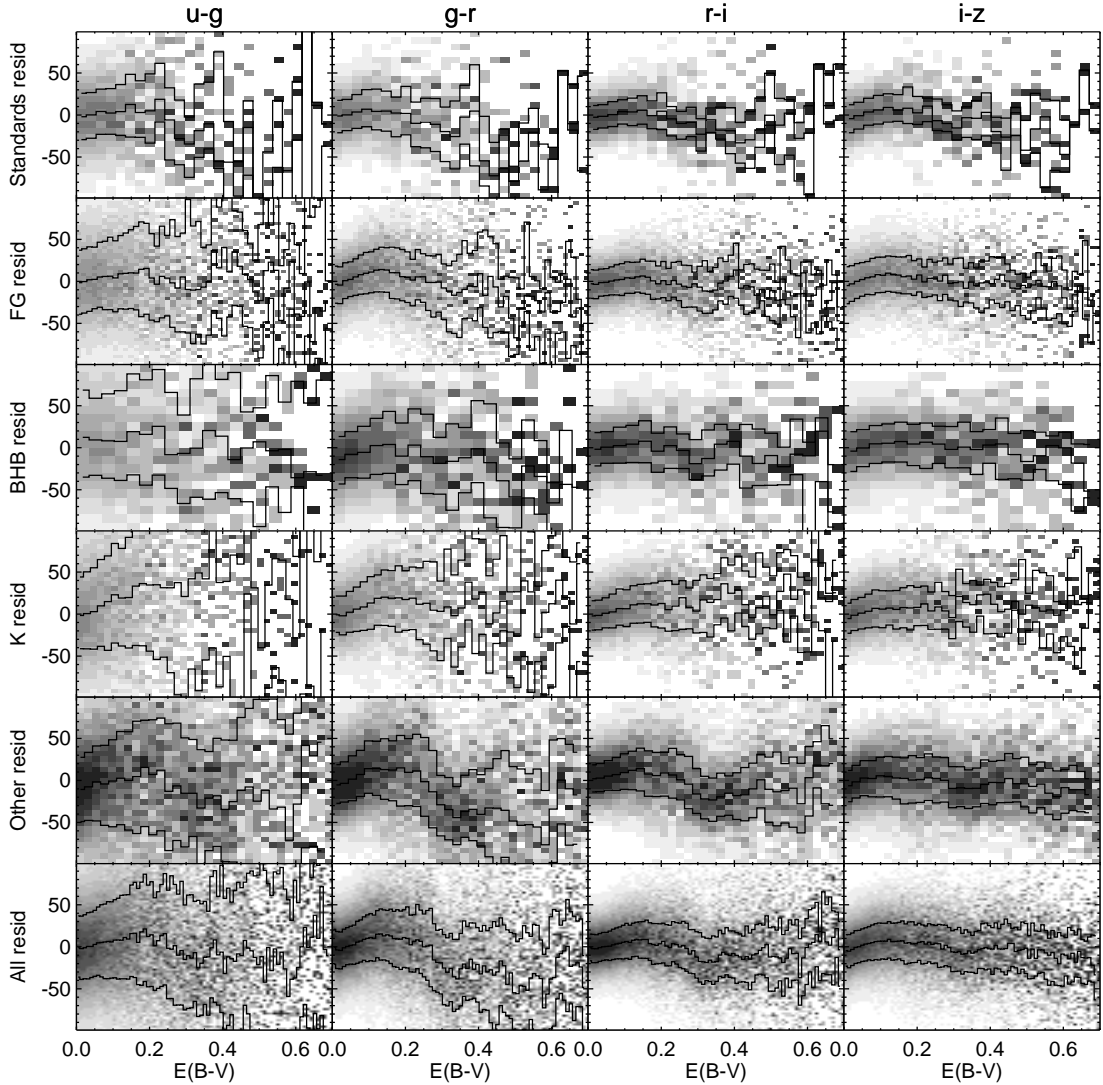

\dfplot{extresid2.ps}
\figcaption{
\label{fig:extresid}
Residuals $\chi$ (Equation \ref{eq:2}) in mmag for the fit without run offsets, as a function of $E(B-V)$ in mag.  The four columns give the colors $u-g$, $g-r$, $r-i$, and $i-z$.  The rows show the residuals for different types of stars: the standard stars, FG-type stars, BHB-type stars, K-type stars, other stars, and all of the stars used in this work.  There is a consistent trend in the residuals to increase when $E(B-V) \lesssim 0.2$, and to decrease thereafter.
}
\end{figure*}

The residuals as a function of $\EBVSFD$ display a striking trend, shown in Figure~\ref{fig:extresid}.  The residuals increase with $E(B-V)$ for $E(B-V) \lesssim 0.2$, and then decrease thereafter.  This result is consistent with the S10 result that the best fit normalization of the dust was larger for $E(B-V) < 0.2$ than it was for $E(B-V) > 0.2$ by about 15\%, which is about the size of the trend seen here.  That said, the entire $E(B-V) \gtrsim 0.3$ footprint for this analysis occurs in a relatively small area near the Galactic anticenter and a few SEGUE low latitude fields, and may not apply more generally.

The formal statistical uncertainties in the fit results are small ($\sim 2$ parts in a thousand).  These uncertainties dramatically underestimate the true uncertainty in the fit, stemming from biases in the photometry (\textsection \ref{subsubsec:photobias}), unmodeled calibration errors in the SDSS, and problems with using SFD as a template (S10).  The reported uncertainties in Table~\ref{tab:fitebv} are derived by a Monte Carlo simulation with simple models for the calibration errors and SFD template errors.  Calibration errors in the SDSS are modeled as normally distributed run zero point errors, with standard deviations of 20 mmags in $u$ and 10 mmags in $g$, $r$, $i$, and $z$, according to the estimated calibration uncertainties of \citet{Padmanabhan:2008}.  We additionally allow the true zero point to change by $\dot{a}t$ over the course of the night, where $\dot{a}$ gives the rate of change of zero point and $t$ is the time of night.  We choose $\dot{a}$ to be normally distributed with standard deviations of 20 mmags per 6 hours in $u$ and 10 mmags per 6 hours in $g$, $r$, $i$, and $z$.  For the SFD template errors, we multiply SFD by a normalization which varies over the sky on 1 degree scales, with a mean of 1 and standard deviation of 0.1.  This is very roughly intended to simulate the uncertainty in the SFD dust temperature correction, but does not include the covariance between temperature and extinction that may be important.  

The resulting uncertainties are dominated by uncertainties coming from the calibration effects, with the zero point and change in zero point with time effects contributing similarly.

\subsection{Fitting Ratios of $R_{a-b}$ using $\D$ Color-Color Diagrams}
\label{subsec:rabcc}

We can free the analysis from a dependence on SFD and its normalization by looking only at ratios of $R_{a-b}$.  If $\D$ has been appropriately calibrated, then in the absence of dust $\D$ should be consistent with zero.  The presence of dust will spread $\D$ along a line in color-color space.  The slope of this line in a pair of colors $a-b$ and $c-d$ gives the ratio $R_{a-b}/R_{c-d}$ and is determined by the reddening law.

\begin{figure*}[tbh]
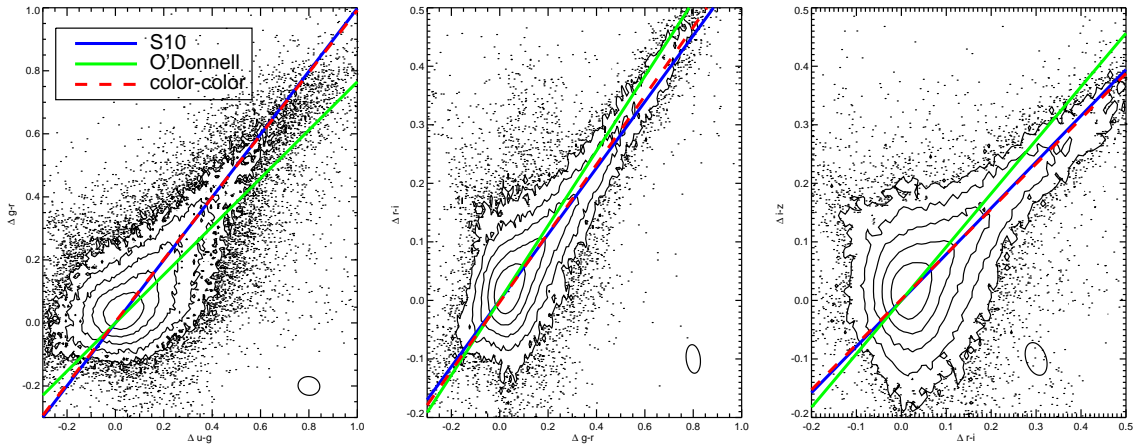

\dfplot{fitccall2.ps}
\figcaption{
\label{fig:fitcc}
Color-color plots of $\D$ in mag.  The data should have $\D$ consistent with zero in the absence of dust.  Reddening from dust spreads $\D$ along a line given by the reddening law.  The predictions for this line for the \citet[S10]{Schlafly:2010} measurements and for an \citet{O'Donnell:1994} reddening law are given by the blue and green lines.  The red line is the best fit to the data, and is consistent with the S10 measurements but not with the \citet{O'Donnell:1994} prediction.  The oval in the lower right of the panels gives a typical uncertainty covariance ellipse for a point on the diagram.
}
\end{figure*}

We find the best-fit the slope of the line in $\D$ in color space in a least-squares sense, considering the covariance between the $\D[a-b]$ induced by the photometry and the color predictions from the stellar parameters.  Figure~\ref{fig:fitcc} shows the result of this fit, together with the predicted line from S10 and from an \citet{O'Donnell:1994} reddening law.  Table~\ref{tab:fitcc} tabulates the best fit parameters for the ratios of $R_{a-b}$.  The fits do a good job at reproducing the S10 ratios for $R_{a-b}$, but disagree with the \citet{O'Donnell:1994} reddening law predictions.  We achieve $\chi^2/\mathrm{dof}$ of 1.4 on average over all the stars we fit.  On the more limited subset of standard stars we get $\chi^2/\mathrm{dof}$ of 1.1.

An important caveat applies to this analysis.  The $\D[a-b]$ for a star are correlated.  We track the correlation induced by the uncertainty in the photometry and the uncertainty in the stellar parameters.  However, any unknown effect that changes the predicted or measured colors of stars will additionally spread out $\D$ in color-color space, and there is a danger of interpreting that signal as signal from the dust.  When instead we fit $\D$ as a function of $\EBVSFD$, we have the advantage of using a template that is uncorrelated with the uncertainties in $\D$ and that is known to be tightly correlated with the dust.

\begin{deluxetable}{c c c c}
\tablewidth{\columnwidth}
\tablecaption{Color-color Fit Results}
\tablehead{
\colhead{Color} & \colhead{Color-color fit} & \colhead{S10} & \colhead{O'Donnell}
}
\startdata
$R_{g-r}/R_{u-g}$ &   $0.99 \pm   0.02$ &   1.00 &   0.77 \\ 
$R_{r-i}/R_{g-r}$ &   $0.59 \pm   0.02$ &   0.57 &   0.64 \\ 
$R_{i-z}/R_{r-i}$ &   $0.77 \pm   0.02$ &   0.79 &   0.91
\enddata
\tablecomments{
\label{tab:fitcc}
Ratios of $R_{a-b}$ derived from fitting $\D$ color-color diagrams.  The spectrum-based $R_{a-b}$ are seen to closely match the \citet[S10]{Schlafly:2010} measurement, while disagreeing with the O'Donnell prediction.
}
\end{deluxetable}

To verify the robustness of these results we have separately fit the standard stars, the F and G stars, the BHB stars, the K stars, and all of the target types we look at.  The results were consistent at the 3\% level.

\subsection{Sources of Systematic Error}
\label{subsec:bias}

\subsubsection{The Photometric Bias}
\label{subsubsec:photobias}

Most SDSS stars are targeted for spectra on the basis of their photometry.  Usually stars are targeted for spectra if they fall within a specific box in color space.  Photometric uncertainty may scatter stars into the color box preferentially from one side if the box is placed in a region of color space where the gradient in stellar number density with color is different on one side of the box than the other.  This effect will lead to a bias in the photometry.

Were this bias a constant for the entire survey it would be taken out by the calibration of $\D$.  However, most of the SDSS and SEGUE color boxes select stars on the basis of their SFD-dereddened colors.  Thus, errors in SFD and its normalization can essentially shift the color box relative to the stellar locus and introduce a bias that changes systematically with $\EBVSFD$.

We have analyzed the magnitude of the bias introduced by this effect by simulating the photometric bias for each of the SDSS and SEGUE target types.  We take the SDSS stellar locus at high Galactic latitudes as a proxy for the true stellar locus.  Errors are artificially added to the stars' magnitudes consistent with the SDSS estimates for the uncertainties to create a new observed stellar locus.  The mean error of a star making it into a color box gives the photometric bias.  The change in the photometric bias with $\EBVSFD$ can be investigated by shifting the colors of the stars according to the error in $\EBVSFD$.

For the target types considered in this work, the photometric bias is less than 0.02 mags, though for very red or unusual types it can be much larger.  The potential bias introduced in the derived $R_{a-b}$ is 2\%.

\subsubsection{The Three Dimensional Footprint}
\label{subsubsec:threed}

The stars targeted by the SDSS for spectra probe a range of distances.  For the dwarf stars in the survey, we use the \citet{Juric:2008} ``bright'' photometric parallax relation to get approximate distances to stars from their colors.  We find that the stars we analyze have distances from 1--6 kpc, with a long tail to greater distances.

The SFD dust map was calibrated to match the reddening of galaxies, which will be behind any dust in the Galaxy.  If substantial dust exists beyond the distances probed by the stars in our data, the $R_{a-b}$ found in this work will be less than the true values.  However, we find compatible $R_{a-b}$ for the K stars ($\sim$ 1 kpc away) as for the standard stars in the sample ($\sim$ 4 kpc away), indicating that the dust is confined to within 1 kpc over most of the SDSS footprint.  In future work we intend to study the three-dimensional distribution of the dust by probing nearer distances with redder stars and by focusing on regions of lower Galactic latitude.

\subsection{Fits to Different Sky Regions}
\label{subsec:regions}

\begin{figure*}[tbh]
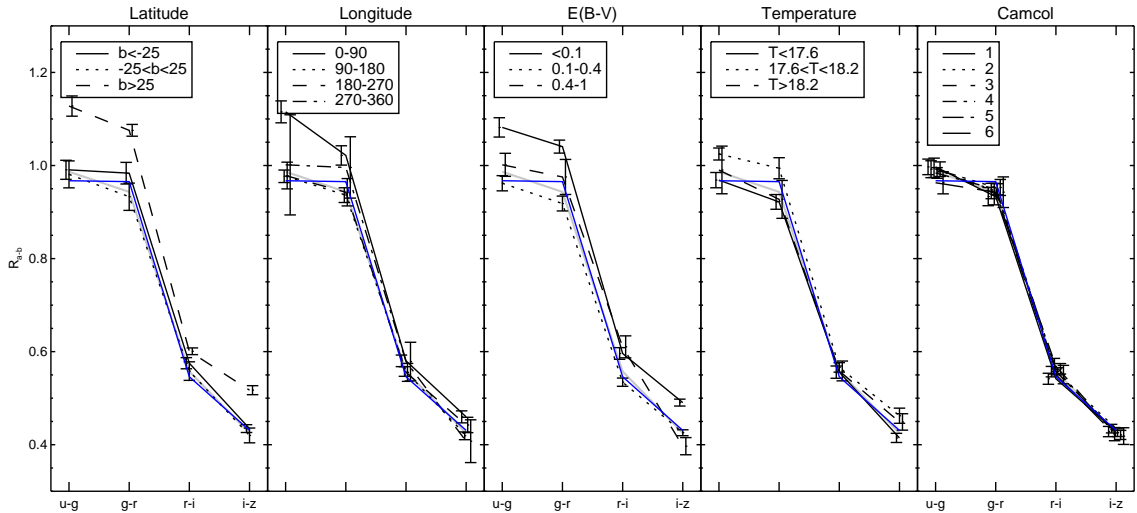

\dfplot{skysubset.zp.ps}
\figcaption{
\label{fig:skysubset.zp}
$R_{a-b}$ for various subsets of the full data for each of the SDSS colors, using fits with run offsets.  Shown are $R_{a-b}$ for the north ($b > 25\degree$), south ($b < -25\degree$), and plane ($|b| < 25\degree$); $R_{a-b}$ for four ranges of Galactic longitude; $R_{a-b}$ for regions of increasing \EBVSFD; $R_{a-b}$ for regions of different temperature; and finally $R_{a-b}$ for the 6 SDSS camera columns.  Overplotted in blue are the \citet{Schlafly:2010} values divided by 1.04 (\textsection \ref{subsec:rab}); the thick gray line gives the fit over all the data.  Error bars are slightly offset horizontally for legibility.  The fitted $R_{a-b}$ are reasonably uniform over the sky, except most noticeably for a 10\% normalization difference between the fit results in the Galactic north and south.
}
\end{figure*}

The $\D$ are sufficiently numerous and cover a sufficiently large region of the sky that they can be usefully cut into different subsets to study the variation in $R_{a-b}$ over the sky.  Figure~\ref{fig:skysubset.zp} shows the results of such an analysis.  The panels show $R_{a-b}$ for different ranges in Galactic latitude, longitude, \EBVSFD, and dust temperature according to SFD.  The rightmost panel shows $R_{a-b}$ as derived for stars observed with each of the SDSS camcols.  All $R_{a-b}$ are fit with zero point offsets to minimize sensitivity to calibration errors in the SDSS.

The fits results show smaller scatter between the various sky regions.  We recover the S10 result that stars with $\EBVSFD < 0.1$ prefer a larger normalization than other stars by $\sim 10\%$; likewise in the north relative to the south, and possibly for the same reason, as the north has less dust than the south.  Fits to the dust in different ranges of Galactic longitude and at different temperatures according to SFD are reasonably consistent.

We can get some idea as to whether or not the filter differences between the SDSS camcols are important to this analysis by splitting the survey by camera column.  We find very close (2\%) agreement between the various camcols.  This also serves to place an upper bound on the statistical uncertainty in the measurements.

These results on the universality of the $R_{a-b}$ are consistent with those found in S10.  In that work, we did a more detailed analysis of the variation in $R_{a-b}$ over the sky and found the dominant mode of variation in $R_{a-b}$ took the form of an overall 10\% variation in the overall normalization of the $R_{a-b}$.  We have not repeated that analysis here, but we try to account for the effects of a 10\% normalization variation in our uncertainties (\textsection \ref{subsec:rab}).

\subsection{North/South Color Asymmetry}
\label{subsec:nsdiff}

In S10, we noted an asymmetry between the north and south in the dereddened color of the main-sequence turn-off, which we called the blue tip color.  We were unable to tell whether this color difference was a calibration effect or a result of the south having an intrinsically different blue tip color than the north.  Because the spectrum-based reddening measurements are sensitive to calibration problems and not to the intrinsic color of the blue tip, we can use them to break this degeneracy.

Table~\ref{tab:nsdiff} shows the differences in $\D$ between the Galactic north and south, compared with the blue tip color differences found in S10.  We find the offset in $\D$ for each target class, and report the mean of these offsets in Table~\ref{tab:nsdiff}.  For the uncertainty in the means, we report the standard deviation of the offsets.  The color differences are comparable in $r-i$ and $i-z$, suggesting that calibration errors in the SDSS are the source of the discrepancy in these colors.  In $g-r$, however, the blue tip color differences are larger than the spectrum-based differences.  In $u-g$, the uncertainty is too large to make any definitive statement.  The difference in these colors could then be caused by a fundamentally different blue tip color in the north than in the south.  This result is heartening because the color of the blue tip varies due to age and metallicity changes in stellar populations, which should primarily affect the $u-g$ and $g-r$ colors while leaving the $r-i$ and $i-z$ colors unchanged.

\begin{deluxetable}{c c c c c}
\tablewidth{\columnwidth}
\tablecaption{Color difference between North and South}
\tablehead{
\colhead{Method} & \colhead{$u-g$} & \colhead{$g-r$} & \colhead{$r-i$} & \colhead{$i-z$}
}
\startdata
Spectrum &   $2.3 \pm 7.2$ &  $8.8 \pm 1.5$ &   $3.4 \pm 1.9$ &   $9.3 \pm 1.6$ \\ 
Blue tip &   7.6 &  21.8 &   7.2 &  12.4 
\enddata
\tablecomments{
\label{tab:nsdiff}
Median color difference in mmag between the north and south, for $40 < |b| < 70$, using the spectrum-based and blue tip methods for measuring colors.  A calibration or dust effect will create a color difference in both sets of measurements, while the presence of a stellar structure in the south would create a color difference only in the blue tip measurements.  The blue tip offsets in $r-i$ and $i-z$ are similar to those in the spectrum-based measurements, suggesting that these color differences do not come from a stellar structure.  However, in $g-r$, the color difference could indicate different stellar populations in the north and south.  In $u-g$, systematic uncertainty in the offsets prevent determination of whether the offset comes from calibration or is astronomical.
}
\end{deluxetable}

\section{Conclusion}
\label{sec:conclusion}

The SDSS provides a wealth of information for testing reddening.  Using the SSPP stellar parameters, our spectrum-based technique achieves reddening measurements with empirical uncertainties of 56, 34, 25, and 29 mmag in the colors $u-g$, $g-r$, $r-i$, and $i-z$, comparable with the expected uncertainties from the photometry and stellar parameters.  Spectra of a few blue stars on a sight line are sufficient to tightly constrain reddening provided that well-calibrated photometry is available and that enough spectra are available to calibrate the predicted colors to the measured colors.  

The use of individual stars with spectra additionally allows extinction to be studied as a function of distance.  Stellar spectra permit good estimates of the intrinsic luminosity of a star in addition to its reddening, making three-dimensional studies of the dust feasible.  The wide range of SEGUE target types---M and K dwarfs through BHB stars---will likely make such analyses especially fruitful, at least at low Galactic latitudes where all of the stars are not behind the entire dust column.  At high Galactic latitudes, we could in principle use large numbers of stars to test for reddening through intermediate and high-velocity clouds.  We defer to later work the attempt to extend this analysis into the third dimension, having established the feasibility and accuracy of the method in two dimensions.  The combination of large photometric and spectroscopic surveys like Pan-STARRS \citep{Kaiser:2002} and LAMOST \citep{Su:1998} seems particularly promising.

These tests should also provide useful feedback to models of synthetic spectra and to stellar parameter estimates.  We look forward to incorporating the DR8 version of the SSPP into this analysis, and potentially pairing it with alternative synthetic spectral grids.  However, we note that while this analysis provides an effective test of the colors predicted by synthetic spectral grids, nevertheless it does not depend on the accuracy of these colors.  The wide range of spectral types targeted by the SDSS permits the predicted colors to be well calibrated to the observed colors, rendering the final results independent of the color accuracy of the synthetic spectra.

The spectrum-based reddening measurements give best-fit $R_{a-b}$ that closely agree with the S10 values.  Accordingly, we have gained confidence in the F99 reddening law with $R_V = 3.1$ and normalization $N = 0.78$ proposed in S10.  The variation in the best-fit normalization of the reddening law seen in that analysis remains a problem.  The spectrum-based reddening tests permit the variation in best-fit normalization with extinction to be seen more clearly; when $\EBVSFD \lesssim 0.2$, the best-fit normalization is about 15\% higher than when $E(B-V) \gtrsim 0.2$, though this conclusion relies on the relatively small fraction of the sky where $E(B-V) \gtrsim 0.3$ and SDSS data is available.

Nevertheless, the agreement of the S10 and spectrum-based reddening measurements demands that the F99 reddening law be used to predict reddening over \citet{Cardelli:1989} or \citet{O'Donnell:1994} reddening laws.  Appendix~\ref{app:ext} gives the extinction per unit $\EBVSFD$ predicted by these sets of measurements in 88 bandpasses for 4 values of $R_V$.  We have focused on high Galactic latitudes where we observe $R_V=3.1$; other values of $R_V$ are provided only for convenience.  Because we have seen that the best fit reddening law normalization varies over the sky and as a function of $\EBVSFD$, it is possible that outside the SDSS footprint a different normalization might be preferable.  However, the shape of the reddening law seems constant over the SDSS footprint, and the normalization we suggest is unambiguously the best choice over the large area of sky covered by the SDSS.  Therefore we propose that this reddening law and normalization become the default choice to be used in the absence of other information.

Multiple sensitive, mutually consistent measurements of reddening over the SDSS footprint are now available.  Extension of this work to larger areas of sky seems readily possible as Pan-STARRS and LAMOST data become available.  These measurements will permit a next generation dust map to be constructed and tested.  We defer to future work the construction of a new dust map that incorporates the insight gained from these measurements.

David Schlegel suggested using SDSS stellar spectra to measure reddening several years ago, and his encouragement and advice were invaluable at the inception of this project.  D.F. and E.S. acknowledge support of NASA grant NNX10AD69G for this research.

Funding for SDSS-III has been provided by the Alfred P. Sloan Foundation, the Participating Institutions, the National Science Foundation, and the U.S. Department of Energy. The SDSS-III web site is http://www.sdss3.org/.

SDSS-III is managed by the Astrophysical Research Consortium for the Participating Institutions of the SDSS-III Collaboration including the University of Arizona, the Brazilian Participation Group, Brookhaven National Laboratory, University of Cambridge, University of Florida, the French Participation Group, the German Participation Group, the Instituto de Astrofisica de Canarias, the Michigan State/Notre Dame/JINA Participation Group, Johns Hopkins University, Lawrence Berkeley National Laboratory, Max Planck Institute for Astrophysics, New Mexico State University, New York University, Ohio State University, Pennsylvania State University, University of Portsmouth, Princeton University, the Spanish Participation Group, University of Tokyo, University of Utah, Vanderbilt University, University of Virginia, University of Washington, and Yale University.

\ 


\appendix
\label{app:ext}
\section{Extinction in Different Bandpasses}

The spectrum-based reddening measurements, in concert with these and the \citet[S10]{Schlafly:2010} reddening measurements, demand that the colors of high latitude stars with $E(B-V) < 1$ be dereddened with an $R_V=3.1$ F99 reddening law rather than an \citet{O'Donnell:1994} reddening law.  The F99 predictions for $R_{a-b}$ agree with the blue tip values to within 3\% in $g-r$, $r-i$, and $i-z$, and are off by 6\% in $u-g$.  This agreement is sufficiently good that we have confidence that reddening in non-SDSS bands can reliably be predicted according to the F99 reddening law.  Accordingly, we have tabulated the reddening per unit $\EBVSFD$ for a large number of commonly used bandpasses (Table~\ref{tab:extbandpass}).

For this computation, following the notation of SFD, we compute
\begin{equation}
\Delta m_b = -2.5 \log \left[ \frac{\int d\lambda W_b(\lambda) S(\lambda)10^{-A(\lambda)\Delta m_{1\mathrm{\mu m}}/2.5}}{\int d\lambda W_b(\lambda)S(\lambda)} \right]
\end{equation}
where $W$ is the system throughput for the band $b$, the source spectrum is $S$ in photons/\AA/s, $A$ is the extinction law, normalized so $A_{1\mum} = 1$, and $\Delta m_{1\mum}$ is $N$ times the extinction at $1 \mum$ according to SFD, following the $R_V = 3.1$ O'Donnell extinction law assumed by SFD.  The factor $N$ here serves as a normalization factor; we use $N = 0.78$ following the results of S10.

The extinction law is parameterized by $R_V = A_V/E(B-V)$.  We have tabulated $\Delta m_b/\EBVSFD$ for $R_V = 2.1$, 3.1, 4.1, and 5.1, in the limit that $\EBVSFD$ is small.  The traditional value of $R_V$ in the diffuse interstellar medium is 3.1 \citep{Cardelli:1989}, which is also what we observe in this work.  Values of $R_V$ other than 3.1 are included in the table only for convenience.

For the source spectrum, we use for $S(\lambda)$ a synthetic spectrum from \citet{Munari:2005}, with $\Teff = 7000 \K$, $\logZ = -1$, and $\logg = 4.5$.

We tabulate values of $\Delta m_b/\EBVSFD$ for the filters considered by SFD, using the F99 reddening law.  For the Landolt $UBVR_{KC}I_{KC}$, CTIO $UBVR_{KC}I_{KC}$, Str\"{o}mgren $ubv\beta y$, Gunn $griz$, Spinrad $R_S$, UKIRT $JHKL^\prime$, and HST WFPC2 bandpasses we follow the procedure detailed in the SFD appendix, combining the filter response with the optical and detector throughput as well as the atmospheric transparency at the usual site.  The SFD appendix used estimated filters for the SDSS from 1994; we now use updated curves taken in June 2001.

The values of $\Delta m_b/\EBVSFD$ incorporate a few other changes relative to those in SFD.  The CTIO atmosphere model SFD used cut off at $3200 \AAm$, slightly affecting the predicted $\Delta m_b/\EBVSFD$ in blue bands using that model.  The source spectrum was also spuriously multiplied by a factor of $\lambda$, for $W_b$ for which SFD did not use a separate detector efficiency curve (UKIRT, Gunn, SDSS, WFPC2, and DSS filter systems).  These shortcomings have been corrected.

We additionally tabulate $\Delta m_b/\EBVSFD$ for new filter sets that have come to prominence since 1998.  We have added values of $\Delta m_b/\EBVSFD$ for the PS1 filter complement, using the throughput measurements from \citet{Stubbs:2010}, and for the LSST target filter complement, using the ``baseline'' expected throughput, downloaded from the LSST public SVN repository \citep{Tyson:2002}.  Finally, we have computed $\Delta m_b/\EBVSFD$ for the wide pass filters available to the Hubble Space Telescope's WFC3 and ACS detectors, using the STSDAS package.  STSDAS is a product of the Space Telescope Science Institute, which is operated by AURA for NASA.

The \citet{Munari:2005} synthetic spectra cover the wavelength range $2500\AAm-10500\AAm$.  This means that values of $\Delta m_b/\EBVSFD$ for bandpasses that cover regions outside this range use extrapolated source spectra.  We extrapolate into the infrared following a blackbody spectrum ($S(\lambda) \propto \lambda^{-3})$, which is more blue than the typical spectrum and so biases the $\Delta m_b/\EBVSFD$ somewhat high.  This affects the UKIRT $JHKL^{\prime}$ and HST F105W, F110W, F125W, F140W, and F160W filters.  The effect is small, however: compared to an $S(\lambda) \propto \lambda^{-1}$ spectrum, $\Delta m_b/\EBVSFD$ is about 1\% larger for the UKIRT filters.  

The CTIO atmospheric transparency spectrum used for the CTIO and Landolt filters cuts off in the blue at 3200\AA.  Blueward of this cutoff we extrapolate that the transmission is constant, rendering the transmitted spectrum too blue and deriving $\Delta m_b/\EBVSFD$ somewhat too large.  This affects only the CTIO and Landolt $U$ bands.


\begin{deluxetable}{c c|c c c c||c c|c c c c}
\tablewidth{\columnwidth}
\tablecaption{F99 Reddening in Different Bandpasses}
\tablehead{
\multirow{2}{*}{Bandpass} & \multirow{2}{*}{$\lambda_\mathrm{eff}$} & \multicolumn{4}{c||}{$R_V$} & \multirow{2}{*}{Bandpass} & \multirow{2}{*}{$\lambda_\mathrm{eff}$} & \multicolumn{4}{c}{$R_V$} \\
\multicolumn{2}{c|}{ } & 2.1 & 3.1 & 4.1 & 5.1 & \multicolumn{2}{c|}{ } & 2.1 & 3.1 & 4.1 & 5.1
}
\startdata

           Landolt $U$ &    3508.2 &   5.614 &   4.334 &   3.773 &   3.460 &            WFPC2 F300W &    3087.6 &   6.777 &   4.902 &   4.127 &   3.710 \\ 
           Landolt $B$ &    4329.0 &   4.355 &   3.626 &   3.290 &   3.096 &            WFPC2 F450W &    4587.0 &   4.014 &   3.410 &   3.132 &   2.971 \\ 
           Landolt $V$ &    5421.7 &   2.953 &   2.742 &   2.645 &   2.589 &            WFPC2 F555W &    5439.4 &   2.976 &   2.755 &   2.653 &   2.594 \\ 
           Landolt $R$ &    6427.8 &   2.124 &   2.169 &   2.189 &   2.201 &            WFPC2 F606W &    5984.8 &   2.469 &   2.415 &   2.389 &   2.375 \\ 
           Landolt $I$ &    8048.4 &   1.410 &   1.505 &   1.548 &   1.573 &            WFPC2 F702W &    6887.9 &   1.850 &   1.948 &   1.994 &   2.020 \\ 
              CTIO $U$ &    3733.9 &   5.170 &   4.107 &   3.628 &   3.355 &            WFPC2 F814W &    7940.0 &   1.452 &   1.549 &   1.594 &   1.620 \\ 
              CTIO $B$ &    4308.9 &   4.382 &   3.641 &   3.300 &   3.104 &             WFC3 F105W &   10438.9 &   0.981 &   0.969 &   0.964 &   0.961 \\ 
              CTIO $V$ &    5516.6 &   2.857 &   2.682 &   2.600 &   2.553 &             WFC3 F110W &   11169.7 &   0.907 &   0.881 &   0.870 &   0.863 \\ 
              CTIO $R$ &    6520.2 &   2.055 &   2.119 &   2.149 &   2.166 &             WFC3 F125W &   12335.5 &   0.778 &   0.726 &   0.701 &   0.687 \\ 
              CTIO $I$ &    8006.9 &   1.420 &   1.516 &   1.561 &   1.587 &             WFC3 F140W &   13692.3 &   0.672 &   0.613 &   0.586 &   0.570 \\ 
             UKIRT $J$ &   12482.9 &   0.764 &   0.709 &   0.684 &   0.669 &             WFC3 F160W &   15258.3 &   0.570 &   0.512 &   0.485 &   0.469 \\ 
             UKIRT $H$ &   16588.4 &   0.502 &   0.449 &   0.425 &   0.411 &            WFC3 F200LP &    5515.2 &   3.457 &   2.958 &   2.743 &   2.625 \\ 
             UKIRT $K$ &   21897.7 &   0.331 &   0.302 &   0.288 &   0.280 &             WFC3 F218W &    2248.3 &  12.405 &   7.760 &   5.956 &   5.027 \\ 
      UKIRT $L^\prime$ &   37772.5 &   0.159 &   0.153 &   0.150 &   0.148 &             WFC3 F225W &    2394.0 &  10.907 &   6.989 &   5.458 &   4.666 \\ 
              Gunn $g$ &    5200.0 &   3.225 &   2.914 &   2.770 &   2.687 &             WFC3 F275W &    2742.5 &   7.986 &   5.487 &   4.488 &   3.963 \\ 
              Gunn $r$ &    6628.5 &   1.959 &   2.055 &   2.099 &   2.125 &             WFC3 F300X &    2934.5 &   7.437 &   5.228 &   4.331 &   3.854 \\ 
              Gunn $i$ &    7898.6 &   1.454 &   1.555 &   1.601 &   1.628 &             WFC3 F336W &    3366.4 &   5.835 &   4.453 &   3.853 &   3.519 \\ 
              Gunn $z$ &    9050.1 &   1.188 &   1.234 &   1.255 &   1.267 &            WFC3 F350LP &    5877.1 &   2.876 &   2.624 &   2.509 &   2.443 \\ 
         Spinrad $R_S$ &    6927.3 &   1.810 &   1.921 &   1.972 &   2.002 &             WFC3 F390W &    3994.8 &   4.803 &   3.896 &   3.481 &   3.244 \\ 
     Str\"{o}mgren $u$ &    3510.0 &   5.539 &   4.305 &   3.759 &   3.452 &             WFC3 F438W &    4335.3 &   4.347 &   3.623 &   3.288 &   3.095 \\ 
     Str\"{o}mgren $b$ &    4670.5 &   3.916 &   3.350 &   3.089 &   2.938 &             WFC3 F475W &    4785.0 &   3.755 &   3.248 &   3.013 &   2.878 \\ 
     Str\"{o}mgren $v$ &    4119.2 &   4.619 &   3.793 &   3.411 &   3.191 &             WFC3 F475X &    4969.7 &   3.548 &   3.116 &   2.917 &   2.803 \\ 
 Str\"{o}mgren $\beta$ &    4861.3 &   3.655 &   3.183 &   2.966 &   2.840 &             WFC3 F555W &    5302.8 &   3.135 &   2.855 &   2.726 &   2.652 \\ 
     Str\"{o}mgren $y$ &    5478.9 &   2.862 &   2.686 &   2.605 &   2.557 &            WFC3 F600LP &    7362.3 &   1.688 &   1.781 &   1.824 &   1.849 \\ 
              SDSS $u$ &    3586.8 &   5.419 &   4.239 &   3.715 &   3.419 &             WFC3 F606W &    5868.5 &   2.581 &   2.488 &   2.445 &   2.421 \\ 
              SDSS $g$ &    4716.7 &   3.843 &   3.303 &   3.054 &   2.910 &             WFC3 F625W &    6225.8 &   2.230 &   2.259 &   2.273 &   2.281 \\ 
              SDSS $r$ &    6165.1 &   2.255 &   2.285 &   2.300 &   2.308 &             WFC3 F775W &    7630.9 &   1.533 &   1.643 &   1.694 &   1.724 \\ 
              SDSS $i$ &    7475.9 &   1.583 &   1.698 &   1.751 &   1.782 &             WFC3 F814W &    7983.1 &   1.441 &   1.536 &   1.580 &   1.605 \\ 
              SDSS $z$ &    8922.9 &   1.211 &   1.263 &   1.286 &   1.300 &            WFC3 F850LP &    9149.7 &   1.168 &   1.208 &   1.226 &   1.237 \\ 
            DSS-II $g$ &    4620.6 &   3.970 &   3.381 &   3.110 &   2.954 &              ACS clear &    6211.1 &   2.612 &   2.436 &   2.356 &   2.309 \\ 
            DSS-II $r$ &    6545.5 &   1.991 &   2.088 &   2.133 &   2.159 &              ACS F435W &    4348.3 &   4.330 &   3.610 &   3.278 &   3.087 \\ 
            DSS-II $i$ &    8111.0 &   1.396 &   1.487 &   1.530 &   1.554 &              ACS F475W &    4760.3 &   3.787 &   3.268 &   3.028 &   2.890 \\ 
               PS1 $g$ &    4876.7 &   3.634 &   3.172 &   2.958 &   2.835 &              ACS F550M &    5581.0 &   2.754 &   2.620 &   2.558 &   2.522 \\ 
               PS1 $r$ &    6200.1 &   2.241 &   2.271 &   2.284 &   2.292 &              ACS F555W &    5361.3 &   3.031 &   2.792 &   2.682 &   2.618 \\ 
               PS1 $i$ &    7520.8 &   1.568 &   1.682 &   1.734 &   1.765 &              ACS F606W &    5901.0 &   2.555 &   2.471 &   2.431 &   2.409 \\ 
               PS1 $z$ &    8665.3 &   1.258 &   1.322 &   1.352 &   1.369 &              ACS F625W &    6298.1 &   2.171 &   2.219 &   2.241 &   2.254 \\ 
               PS1 $y$ &    9706.3 &   1.074 &   1.087 &   1.094 &   1.097 &              ACS F775W &    7673.5 &   1.520 &   1.629 &   1.679 &   1.708 \\ 
               PS1 $w$ &    6240.8 &   2.425 &   2.341 &   2.302 &   2.280 &              ACS F814W &    8012.4 &   1.432 &   1.526 &   1.569 &   1.594 \\ 
              LSST $u$ &    3693.2 &   5.243 &   4.145 &   3.652 &   3.373 &             ACS F850LP &    9007.5 &   1.196 &   1.243 &   1.265 &   1.277 \\ 
              LSST $g$ &    4797.3 &   3.739 &   3.237 &   3.006 &   2.872 &                DES $g$ &    4796.6 &   3.739 &   3.237 &   3.006 &   2.872 \\
              LSST $r$ &    6195.8 &   2.245 &   2.273 &   2.286 &   2.294 &                DES $r$ &    6382.6 &   2.113 &   2.176 &   2.205 &   2.221 \\
              LSST $i$ &    7515.3 &   1.571 &   1.684 &   1.737 &   1.767 &                DES $i$ &    7769.0 &   1.490 &   1.595 &   1.644 &   1.672 \\
              LSST $z$ &    8664.4 &   1.259 &   1.323 &   1.353 &   1.370 &                DES $z$ &    9108.2 &   1.175 &   1.217 &   1.236 &   1.247 \\
              LSST $y$ &    9710.3 &   1.075 &   1.088 &   1.094 &   1.098 &                DES $Y$ &    9850.4 &   1.051 &   1.058 &   1.061 &   1.063   

\enddata
\tablecomments{
\label{tab:extbandpass}
$A_b/\EBVSFD$ in different bandpasses $b$, evaluated according to an F99 reddening law with normalization $N = 0.78$ and $R_V = 2.1$, $3.1$, $4.1$, and $5.1$, using a 7000 K source spectrum.  The column $\lambda_\mathrm{eff}$ gives the throughput-weighted mean wavelength in the bandpass.  When used with $R_V=3.1$, these give the coefficients to use with $\EBVSFD$ to get reddenings consistent with the results of this work and \citet{Schlafly:2010}.  The values for other $R_V$ are provided only for convenience.
}
\end{deluxetable}

\bibliographystyle{apsrmp} 
\bibliography{specdust}

\end{document}